\documentclass[twocolumn]{aastex631}

\begin{document}

\title{Diagnosing the AGN origin of diffuse TeV neutrinos with their spatial correlation}

\author[0000-0003-4907-6666]{Xiao-Bin Chen}
\affiliation{School of Astronomy and Space Science, Nanjing University, Nanjing 210023, China}
\affiliation{Key laboratory of Modern Astronomy and Astrophysics (Nanjing University), Ministry of Education, Nanjing 210023, China}

\author[0000-0003-1576-0961]{Ruo-Yu Liu}
\affiliation{School of Astronomy and Space Science, Nanjing University, Nanjing 210023, China}
\affiliation{Key laboratory of Modern Astronomy and Astrophysics (Nanjing University), Ministry of Education, Nanjing 210023, China}
\affiliation{Tianfu Cosmic Ray Research Center, Chengdu 610000, Sichuan, China}

\author[0000-0002-5881-335X]{Xiang-Yu Wang}
\affiliation{School of Astronomy and Space Science, Nanjing University, Nanjing 210023, China}
\affiliation{Key laboratory of Modern Astronomy and Astrophysics (Nanjing University), Ministry of Education, Nanjing 210023, China}

\correspondingauthor{Ruo-Yu Liu; Xiang-Yu Wang}
\email{ryliu@nju.edu.cn; xywang@nju.edu.cn}

\begin{abstract}
The recent  detection of TeV neutrinos from nearby Seyfert galaxies (e.g., NGC~1068) by IceCube suggests that active galactic nuclei (AGNs) could make a significant contribution to diffuse astrophysical neutrinos. The absence of TeV gamma-rays from NGC~1068 indicates neutrino production in a compact opaque region of gamma-rays. The vicinity of the supermassive black hole, such as the disk-corona, is an ideal region, where the high radiation density leads to efficient neutrino production and  gamma-ray attenuation. Disk-corona models predict that the neutrino emission from AGNs correlates with X-ray emission, which traces the coronal activity. In this paper, we assess whether diffuse TeV neutrinos can originate from X-ray-emitting AGNs with Monte Carlo simulations, considering the predicted performance of future neutrino telescopes. We test this hypothesis by searching for spatial correlations between the X-ray AGN population and high-energy neutrinos, assuming that the neutrino flux scales with the AGN X-ray flux. 
{After accounting for the fraction of the AGN X-ray flux resolved by eFEDS and the remaining unresolved AGN component, we find that an AGN origin of diffuse neutrinos can be tested at a significance of about $3\sigma$ with ten years of IceCube-Gen2 observations. With improved angular resolution and sensitivity, a 30\,km$^{3}$-scale underwater neutrino telescope such as HUNT is expected to reach a significance of about $6\sigma$ with one year of exposure.}
The detection significance decreases if AGNs contribute  partially to the total astrophysical neutrino flux. Our results highlight the critical role of angular resolution in diagnosing the AGN origin of diffuse TeV neutrinos.
 
\end{abstract}


\section{Introduction} 

The IceCube neutrino telescope has detected a significant number of neutrinos with energies between a few TeV and a few PeV. 
Searches for point-like sources associated with known astrophysical objects have largely returned null detection \citep{IC2020prl-ps}, suggesting that the high-energy neutrino sky is dominated by an isotropic background. Although the origins of the vast majority of high-energy neutrinos have not been identified, a significant excess coming from the close-by Seyfert galaxy NGC 1068 at 4.2$\sigma$ level \citep{NGC1068_ic_2022Sci} has been reported recently. Tentative excesses at $3\sigma$ level have also been reported from some other NGC~1068--like AGNs such as  NGC~7469  and NGC~4151\citep{2025arXiv251013403A}, implying that Seyfert galaxies could be the dominant sources of high-energy neutrinos.

The neutrino flux measured from NGC~1068 is more than an order of magnitude larger than the upper limits of TeV gamma-ray emission placed by MAGIC and HAWC \citep{MAGIC1068-2019, HAWC1068-2022}. 
The difference between the observed neutrino flux and gamma-ray flux from NGC~1068 suggests that the neutrinos must be produced in a region that is opaque to GeV–TeV gamma-rays that would otherwise accompany the neutrinos. 
In Seyfert galaxies like NGC 1068, a hot and  magnetized  corona above the disk is formed (see e.g. \citealt{MillerStone2000accretion}). The dense environments near the supermassive black holes together with the acceleration of cosmic rays (CRs) in the corona offer suitable conditions for the production of high energy neutrinos. 
Such a scenario, commonly referred to as disk-corona models, has been shown to be able to accommodate the high level of neutrino emission from the X-ray bright AGN NGC~1068 \citep{Murase2020, Inoue2020, Kheirandish2021, Eichmann2022}.

The large flux contrast between neutrinos and gamma rays in NGC~1068 is in agreement with the picture of gamma-ray ``hidden'' sources for  the origin of the isotropic neutrino background \citep{Senno2015,Murase2016PhRvL,Bechtol2017ApJ,Capanema2020,Fang2022}, as required by comparison of isotropic diffusive neutrino flux measured by IceCube \citep{Aartsen2020PhRvL.125l1104A, ICatmneutrino2022} and the isotropic gamma-ray background observed by Fermi-LAT \citep{Ackermann2015ApJDiffuse}. This consistency further supports the main contributors of the neutrino background are ``hidden'' sources. 
Building on this framework, \cite{Fiorillo2025} extended the corona model for NGC~1068 to the diffuse neutrino contribution of AGN coronae, finding that such sources can account for the IceCube diffuse flux in the $\sim 1–100$\,TeV. { This provides theoretical motivation for testing whether AGN coronae can be the main sources of TeV diffuse neutrinos.}

{
Recent IceCube analyses have reported tentative evidence for neutrino emission from nearby X-ray-bright Seyfert galaxies\citep{Abbasi2022PhRvD, IceCube_ESH_Seyfert_South, IceCube_HardXray_AGN, IceCube_Xray_Bright_AGN}, including NGC~1068 and other similar AGNs. These searches are usually based on stacking or collective analyses of the brightest X-ray-selected sources, such as Swift/BAT Seyferts. They provide important evidence that at least some X-ray-bright AGNs can emit neutrinos. However, such bright-source searches do not show whether the all-sky diffuse neutrino background is dominated by the cumulative emission of the entire AGN population. To test the origin of the majority of the all-sky diffuse background, one needs to search for a statistical spatial correlation between high-energy neutrinos and a sufficiently deep X-ray-selected AGN sample.}

{However, as will be demonstrated later, the limited angular resolution and statistics of current IceCube data make it difficult to establish a significant spatial correlation between neutrinos and a more complete population of X-ray AGNs contributing to the diffuse flux.} 
Future neutrino telescopes, such as IceCube-Gen2 \citep{IceCube-gen22021}, KM3NeT \citep{KM3NET2016JPhG}, Huge Underwater high-energy Neutrino Telescope (HUNT) \citep{HUNTproposal2023}, Tropical Deep-sea Neutrino Telescope (TRIDENT) \citep{TRIDENT2023NA}, NEON \citep{NEON2025} and Baikal-GVD \citep{Baikal2022}, will have larger effective volumes and improved angular precision, and are therefore crucial for testing this correlation. 
Motivated by this, in this work, we assess whether the X-ray AGN origin for diffuse TeV neutrinos can be tested by using the spatial correlation between the  X-ray AGN population and neutrino population  with future neutrino telescopes. We will use IceCube-Gen2 and HUNT as two examples of future neutrino telescopes for the analyses.
We assume the neutrino luminosity to be positively related to the X-ray luminosity of the AGN, given that the X-ray luminosity may reflect the energy production rate in the corona, and is also proportional to the target density for photopion production. By combining the AGN X-ray luminosity distribution with the expected performance of future neutrino telescopes, we forecast the prospects for detecting such an AGN-neutrino correlation.

{The rest of this paper is organized as follows. Section~2 describes the simulation and likelihood framework, Section~3 presents forecasts for future neutrino detectors, Section~4 quantifies the main systematic uncertainties, and Section~5 discusses extensions and consistency checks.}

\section{Analysis technique}

\subsection{Simulating Neutrino-AGN Associations}

We evaluate the spatial correlation between high-energy neutrinos and AGNs by comparing how well the observed neutrino directions agree with the expectations under two competing scenarios for their astrophysical origins:
\begin{enumerate}
    \item \textbf{AGN Association Hypothesis:} A subset or all of the detected neutrinos originate from AGNs, with their flux scaling proportionally to the X-ray fluxes of AGNs. In this scenario, the X-ray flux of an AGN serves as the proxy for its neutrino flux, reflecting the case in which AGNs are genuine neutrino emitters.
    \item \textbf{Null Hypothesis (Random Background):} Neutrino events are uncorrelated with AGNs and are isotropically distributed. Any apparent spatial association between AGNs and neutrinos arises purely by coincidence. 
\end{enumerate}

In both models, we account for contamination from atmospheric neutrinos, which form an irreducible background in high-energy neutrino observations. According to the IceCube 9.5-year dataset \citep{ICatmneutrino2022},  the astrophysical signal starts to dominate the atmospheric neutrinos from 100 TeV. This motivates a further decomposition of the observed neutrino sample into its astrophysical and atmospheric components. We introduce the parameter $\beta$, defined as the fraction of astrophysical neutrinos relative to the total neutrino population (including atmospheric background) above a given energy threshold:
\begin{equation}\label{eq:beta}
\beta(E_{\rm th}) \equiv \frac{N_{\nu}^{\rm astro}(E > E_\nu^{\rm th})}{N_{\nu}^{\rm all}(E > E_\nu^{\rm th})}.
\end{equation}
The value of $\beta$ depends sensitively on the chosen neutrino energy threshold $E_\nu^{\rm th}$ and increases rapidly toward unity at higher energies, reflecting the diminishing contribution from atmospheric neutrinos. In this work, $\beta(E_\nu^{\rm th})$ is derived directly from IceCube observations \citep{ICatmneutrino2022}.
In addition, we define $\alpha$ as the fraction of astrophysical neutrinos produced by AGNs:
\begin{equation}\label{eq:alpha}
\alpha \equiv \frac{N_{\nu}^{\rm AGN}} {N_{\nu}^{\rm astro}},
\end{equation}
which characterizes the efficiency of AGNs as neutrino sources relative to the total astrophysical neutrino flux.

The AGN catalog from the eROSITA Final Equatorial-Depth Survey (eFEDS) \citep{eFEDSAGN2022} forms the basis of our analysis. It includes 22,079 AGNs selected by X-rays in a contiguous 142~deg$^2$ field. 
{ In addition to the source detection and photometric information, the eFEDS AGN spectral catalog provides posterior estimates of X-ray spectra, including both observed and absorption-corrected fluxes. In this work, we use the catalog quantity \texttt{FluxcMedT}, defined as the posterior median of the absorption-corrected flux in the $2.3$--$5$~keV band \citep{eFEDSAGN2022}, as the X-ray weight of each AGN. }
Given its depth ($6.5\times 10^{-15} {\rm erg\,s^{-1}\,cm^{-2}}$, \citealt{eFEDS1_2022A&A}) and well-characterized selection, the eFEDS sample provides a deep resolved template of the AGN population accessible to current wide-field X-ray surveys.
{ The limited solid angle of eFEDS also means that it is not expected to contain the rare, nearest, and brightest Seyferts that dominate current IceCube bright-source or stacking signals. This limitation is not in conflict with the goal of the present work, because those brightest X-ray AGNs only constitute a minor fraction of the total X-ray flux from the entire AGN population. }

\begin{figure}
    \centering
    \includegraphics[width=0.48\textwidth]{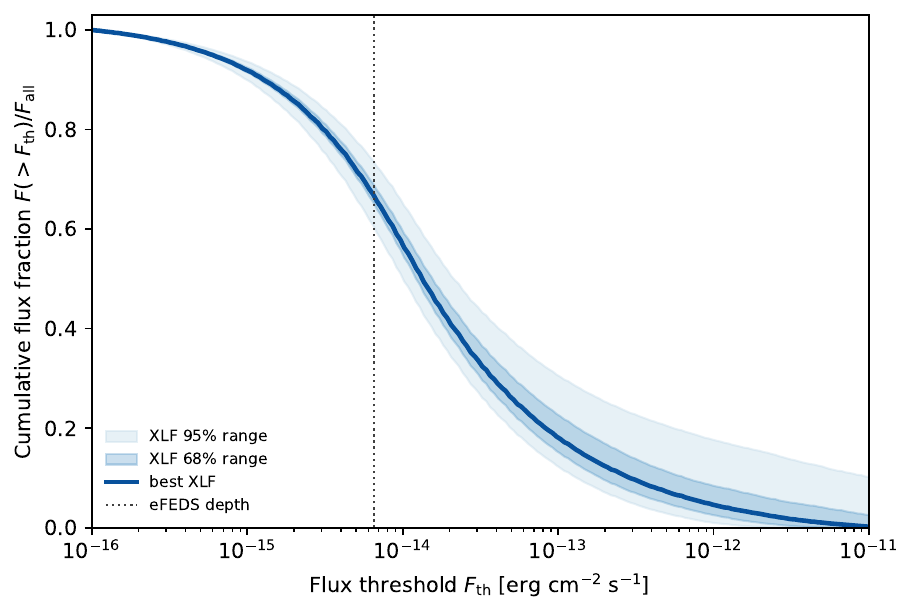}
    \caption{
    {
    XLF-based cumulative AGN flux fraction as a function of the X-ray flux threshold $F_{\rm th}$. The vertical dotted line marks the nominal eFEDS depth and the representative flux cut used in the likelihood analysis.} }
    \label{fig:xlf_cumulative_flux}
\end{figure}

{ To quantify the depth represented by eFEDS, we compare its flux limit with the cumulative AGN flux distribution predicted by the X-ray luminosity function (XLF) of \citet{UedaXLF2014}. As shown in Figure~\ref{fig:xlf_cumulative_flux}, the nominal eFEDS depth corresponds to about $67\%$ of the total XLF-weighted AGN X-ray flux. Therefore, eFEDS is considered as a resolved AGN template for this dominant flux component, rather than as a complete sample of the entire AGN population. The remaining $\sim 33\%$ is attributed to fainter unresolved AGNs below the eFEDS depth. Since these sources are not included in the resolved template, they are treated as an effectively diffuse component in the spatial-correlation simulation.}

Under the AGN-association hypothesis, we normalize the simulation by requiring that the total number of modeling neutrinos equals the expected number derived from the measured diffuse astrophysical neutrino flux plus the atmospheric neutrino background. These fluxes, taken from muon-neutrino flux with 9.5 years of IceCube data \citep{ICatmneutrino2022}, are integrated over the chosen energy threshold and then folded with the effective area of the detector to yield the expected total count of events $N_{\nu}^{\rm tot}$. 
{For a given total AGN contribution $f_{\rm tot}$, only the XLF-weighted fraction represented by the resolved eFEDS template is assigned to cataloged eFEDS sources. At the nominal eFEDS depth, this corresponds to a full resolved-template contribution $f_{\rm eFEDS}\simeq0.67\,f_{\rm tot}$. The remaining unresolved AGN contribution is treated as an isotropic diffuse component.}
The probability of each resolved eFEDS AGN producing a template-associated neutrino is taken to be proportional to its X-ray flux ($F_{X}$): 
\begin{equation} 
P_i \propto F_{X,i}.
\end{equation}
{ This linear relation is adopted as our baseline population-level assumption. Possible source-to-source scatter around this relation is tested separately in Appendix~\ref{app:scatter}. \footnote{Such scatter has been considered in recent IceCube analyses; for example, \citet{Jain2026} reported a $\sim4\sigma$ correlation between X-ray--bright AGNs and IceCube neutrinos after incorporating source-to-source uncertainty in the X-ray--neutrino flux relation into the likelihood function.}}
Because the AGN catalog is flux-limited, the parameter $f$ captures only the contribution from the selected catalog sources ($F_X>F_{\rm min}$). To estimate the total AGN population contribution, including X-ray–faint AGNs below the threshold $F_{\rm min}$, we extrapolate using the AGN XLF. {Here $\sum_{\rm all}F_X$ denotes the full XLF-weighted AGN flux, not the finite eFEDS catalog flux, and $\sum_{F_X>F_{\rm min}}F_X$ denotes the corresponding XLF-weighted flux above the analysis threshold. . For a generic flux threshold $F_{\rm min}$, the total AGN neutrino fraction is then given by
\begin{equation}\label{eq:f_total}
f_{\rm tot} = f \cdot 
\frac{\sum_{\rm all} F_X}{\sum_{F_X>F_{\rm min}} F_X},
\end{equation}}
and the expected value of this quantity is $\alpha \times \beta$.

For each template-associated neutrino assigned, its arrival direction is offset from the source position according to a two-dimensional Gaussian with a width equal to the detector’s point-spread function (PSF).
In contrast, under the null hypothesis, neutrino positions are randomly drawn from a uniform distribution across the same sky footprint, with no relation with AGN locations.

\subsection{Unbinned Likelihood Framework for Spatial Correlation Analysis}\label{sec:llh_analysis}

An unbinned likelihood ratio method can be used to distinguish astrophysical neutrinos from the atmospheric background produced by cosmic ray interactions \citep{BraunML2008, crequeSeekingNeu2022, IC2024MOJAVEAGNApJ}. 
The analysis is designed to measure the fractional contribution of AGN-generated neutrinos to the all-sky neutrino sample.
The optimal estimator to determine the fraction $f$ of neutrinos associated with AGN is obtained using the unbinned maximum likelihood (ML) approach, formulated as
\begin{equation}\label{eq:L_f}
\mathcal{L}(f) = \prod_{i=1}^{N} \left[ f \cdot S_i + (1 - f) \cdot B_i \right],
\end{equation}
where $S_i$ and $B_i$ represent the signal and background probability density functions (PDFs), respectively, for the $i$-th neutrino event.
The signal PDF $ S_i $ represents the likelihood of the $ i $-th neutrino originating from the AGN population, modeled as a weighted sum over the AGNs included in the selected signal template:
\begin{equation}\label{eq:S}
S_i = \sum_j w_j\frac{1}{2\pi\sigma_{\rm psf}^2} \exp\left( -\frac{(\vec{x}_i - \vec{x}_j)^2}{2\sigma_{\rm psf}^2} \right),
\end{equation}
where $ \vec{x}_i $ and $ \vec{x}_j $ are the reconstructed positions of the neutrino and the $j$-th AGN, respectively, and $ w_j = F_j / \sum_j F_j $ reflects the X-ray flux-weighted contribution of each AGN.  
{ Here $F_j$ denotes the absorption-corrected X-ray flux of the $j$-th AGN in the $2.3$--$5$~keV band, as described above.}
The parameter $ \sigma_{\rm psf}$ characterizes the angular resolution (i.e. PSF) of the neutrino detector.
The background PDF $ B_i $ accounts for neutrinos arising from the isotropic atmospheric background and is modeled as a uniform distribution over the analysis region: $B_i = \frac{1}{\Omega}$,
where $ \Omega $ is the solid angle subtended by the sky region covered by the AGN catalog and neutrinos.

The signal fraction $f$ is determined by maximizing the likelihood function $\mathcal{L}(f)$ constructed over all observed neutrino events. {At the likelihood level, $f$ denotes the fraction associated with the AGNs included in the signal template, not the full AGN population.}
Under the null hypothesis (i.e., no AGN contribution), the likelihood reduces to $\mathcal{L}(f=0) = \prod_{i=1}^{N} B_i$. The strength of the signal is then quantified by the likelihood ratio test statistic:
\begin{equation}
    \mathrm{TS} = \lambda = -2 \cdot \log \left[ \frac{\mathcal{L}(f)}{\mathcal{L}(f=0)} \right],
\end{equation}
which measures the relative preference for a nonzero $f$ over the null hypothesis. In the large-sample limit, TS approximately follows a $\chi^2$ distribution with one degree of freedom, and the statistical significance can be estimated as $\sigma \simeq \sqrt{\mathrm{TS}}$. 
The validity of this approximation has been verified with $5\times10^6$ background-only Monte Carlo simulations (see Appendix~\ref{sec:ts_bg}).  

{ Based on the best-fit $f$ from the likelihood method, the total AGN neutrino fraction $f_{\rm tot}$ can be obtained after applying the flux-threshold and XLF population corrections based on Eq.~\ref{eq:f_total}.} 
To quantify the uncertainty on the total AGN neutrino fraction, we define the relative error $\delta$: 
\begin{equation}\label{eq:delta_f}
\delta \equiv \delta_{f_{\rm tot}} = \frac{ \sigma_{f_{\rm tot}} }{f_{\rm tot}},
\end{equation}
where $\sigma_{f_{\rm tot}}$ is obtained from the distribution of $f_{\rm tot}$ over $800$ independent Monte Carlo realizations, and therefore includes both statistical fluctuations (i.e., the Poisson fluctuations in the expected number of neutrino events) and systematic uncertainties, which will be discussed in Section~\ref{sec:uncer}.

\subsection{Mitigating Source Confusion in Analyses} \label{sec:mitigate}

\begin{figure}
    \centering
    \includegraphics[width=0.45\textwidth]{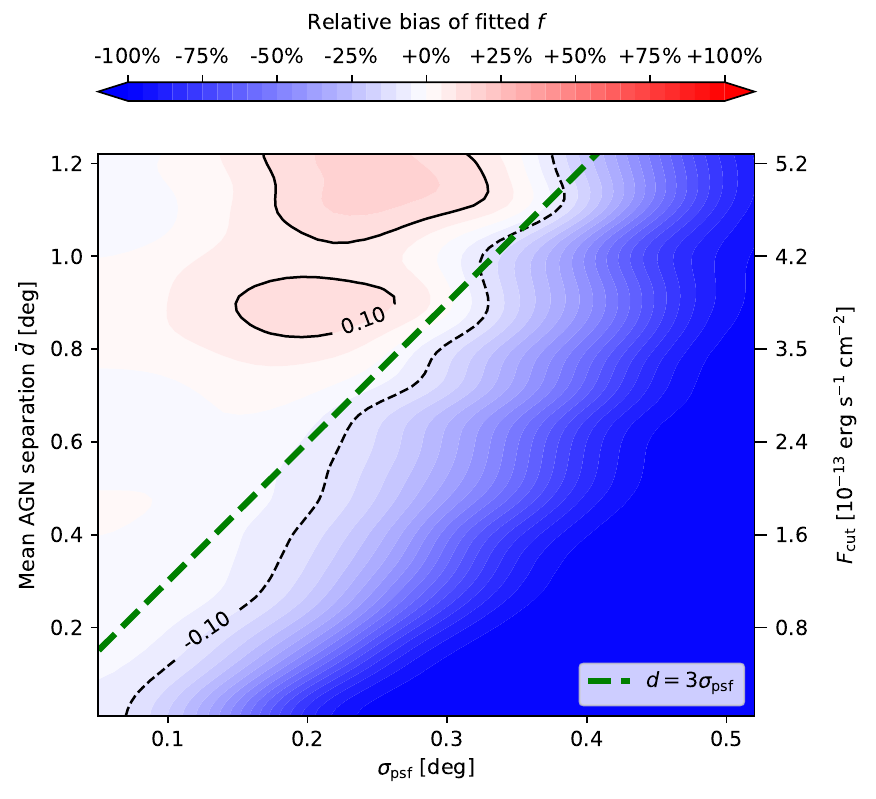}
    \caption{
    {Relative bias of the fitted flux-limited template fraction $f$ as a function of the detector PSF and the average angular separation of the selected AGNs. The bias is defined as the fitted value minus the injected value, divided by the injected value. This diagnostic test is performed within the resolved eFEDS catalog, with the full eFEDS-template contribution fixed to $f_{\rm eFEDS}=0.5$. The right vertical axis gives the corresponding flux cut $F_{\rm cut}$. Larger $\bar d$ corresponds to selecting fewer and brighter AGNs. The green dashed line marks $d=3\sigma_{\rm psf}$. Above this line, the selected-template fraction can be recovered without significant bias.}
    }
    \label{fig:psf_density}
\end{figure}

{Likelihood-based methods provide a useful way to estimate the flux-limited template fraction $f$, which is then converted to the total AGN contribution $f_{\rm tot}$ through the XLF correction. However, this procedure can be biased by source confusion. Source confusion becomes important when the angular resolution is too poor to separate nearby sources }\citep{barcons1992, KurczynskiConfusiong2010,WhenConfusingJones201}. As also noted by \cite{Kowalski2025}, stacking becomes unreliable when the average number of sources within one PSF is larger than unity. { In this regime, the fitted flux-limited  template fraction $f$ and hence of the inferred $f_{\rm tot}$ can be systematically underestimated.}

To quantify this effect, we investigate how the mean angular separation between sources ($\bar{d}$) interacts with the detector's PSF width ($\sigma_{\rm psf}$). 
{ Since brighter sources are more sparsely distributed on the sky, a higher X-ray flux cut reduces the source density and increases $\bar{d}$. In Figure~\ref{fig:psf_density}, the left vertical axis shows $\bar d$, while the right vertical axis shows the corresponding $F_{\rm cut}$ derived from the eFEDS flux distribution. For each detector configuration, We define a flux cut $F_{\rm cut}$, such that the selected AGN subset satisfies $\bar{d}(F_{\rm cut}) \geq 3\sigma_{\rm psf}$.}
This approach allows us to explore a wide range of source densities while maintaining physical realism.

{For the source-confusion diagnostic shown in Figure~\ref{fig:psf_density}, we work only within the resolved eFEDS catalog and fix the full eFEDS-template contribution to $f_{\rm eFEDS}=0.5$. For each source-separation value $\bar d$, the corresponding flux cut $F_{\rm cut}$ selects only a subset of eFEDS AGNs. The figure therefore tests whether this flux-limited $f$ can be recovered before converting it to $f_{\rm tot}$.} 
When the average separation satisfies $\bar{d} \geq 3\sigma_{\rm psf}$, {the likelihood fit recovers the injected selected-template fraction with little bias.} In contrast, for denser source populations ($\bar{d} \leq 3\sigma_{\rm psf}$), confusion effects cause the fitted $f$ to be systematically underestimated.
We note that the present analysis is based on the eFEDS AGN catalog, which covers a relatively limited sky area of $142\,\mathrm{deg}^2$. In regimes where the source density is low and the average separation becomes large, the finite survey area can amplify statistical fluctuations, resulting in increased scatter in the recovered template fraction, as visible in the upper-left region of Figure~\ref{fig:psf_density}. This effect is primarily driven by limited source statistics rather than a failure of the likelihood method itself. In future wide-area X-ray surveys with substantially larger sky coverage, this statistical limitation will be significantly alleviated, leading to more stable and accurate parameter recovery.
This yields a practical condition for method applicability: the source population must be sufficiently sparse, such that the average spacing between sources exceeds approximately three times the PSF width.  Importantly, we verify that this trend is valid for a wide range of values of $f_{\rm eFEDS}$, not just for the specific case of $f_{\rm eFEDS} = 0.5$. 
In the following analyses, this empirical requirement, $\bar{d} \geq 3\sigma_{\rm psf}$, is adopted as a baseline condition to ensure that source confusion does not bias the likelihood inference.

{In the detector applications below, the signal template is further restricted by the source-confusion criterion. We therefore evaluate the generic threshold $F_{\rm min}$ in Eq.~(\ref{eq:f_total}) at the analysis cut $F_{\rm cut}$, where $F_{\rm cut}$ is the value required for the selected AGN subset to satisfy $\bar{d}\geq3\sigma_{\rm psf}$. The likelihood is applied only to AGNs with $F_X>F_{\rm cut}$, and the fitted fraction $f$ represents the neutrino contribution from this selected flux-limited sub-sample.}

\section{Predictive Power for Future Neutrino Detectors}

Next-generation neutrino observatories are expected to greatly enhance the capability to detect astrophysical neutrino sources. Proposed facilities such as IceCube-Gen2, HUNT, TRIDENT, NEON and Baikal-GVD, also promise enhanced exposure and angular resolution. These advances motivate us to examine how the improvements in angular resolution, event statistics, and energy threshold affect the prospects for diagnosing the neutrino–AGN correlation.

In this section, we consider a special scenario in which all astrophysical neutrinos originate from AGNs, corresponding to $\alpha = 1$. This idealized case represents the maximal AGN contribution to the diffuse astrophysical neutrino flux and provides an upper limit on the expected signal in future detectors. Studying this scenario allows us to explore the full potential of next-generation instruments to constrain or detect neutrino–AGN associations under optimal conditions.

We perform a series of simulations to quantify how key observational parameters influence the detection potential. Specifically, we examine four primary factors: the detector’s angular resolution, the total number of detected neutrinos, the minimum energy threshold, and the observation duration. These variables determine both the statistical power and the level of background contamination, which jointly govern the overall sensitivity of correlation searches.

\begin{figure}
    \centering
    \includegraphics[width=0.45\textwidth]{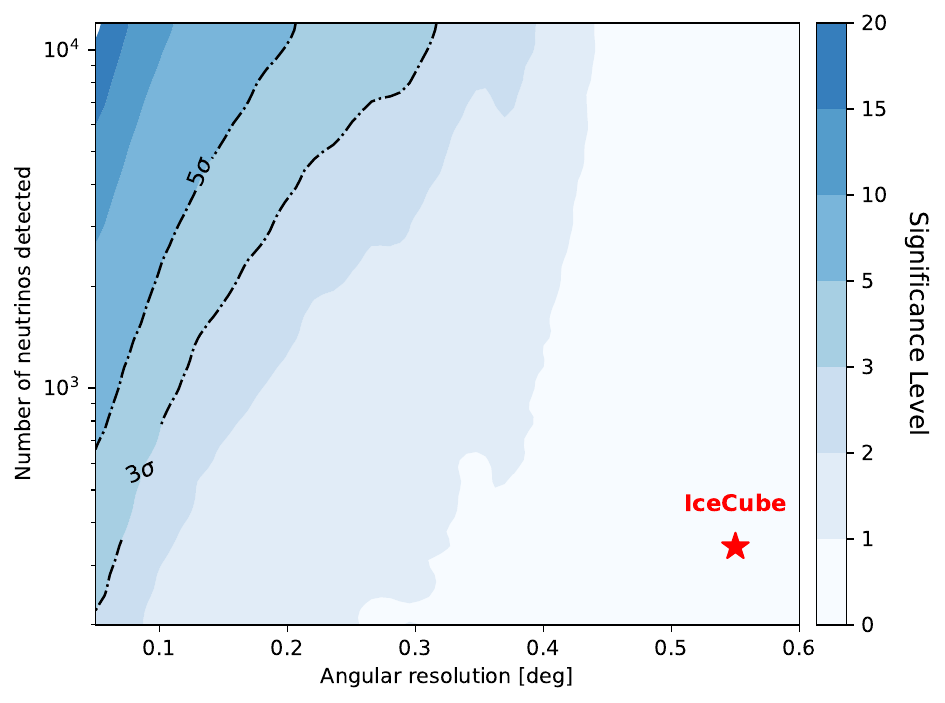}
    \caption{Detection significance as a function of angular resolution and the total number of detected neutrinos. 
    The color map indicates the significance level, with solid and dashed contours marking the 5$\sigma$ and 3$\sigma$ confidence levels, respectively. {The scan assumes a total AGN contribution of $f_{\rm tot}=0.5$, with only the XLF-resolved eFEDS component contributing to the spatial template. In the context of the eFEDS-like population-correlation forecast considered here, the star marker denotes current \textit{IceCube Alert} events \citep{IceCat-2}}. 
   }
    \label{fig:significance}
\end{figure}

Figure~\ref{fig:significance} illustrates how the detection significance depends on the angular resolution and the total number of detected neutrinos. { In this scan, the total AGN-associated fraction is fixed at $f_{\rm tot}=0.5$ and the spatially correlated resolved-template component is $f_{\rm eFEDS}\simeq0.67f_{\rm tot}$. This scan is intended to show the generic dependence on event statistics and angular resolution. The detector-specific forecasts below use the corresponding $\beta$ and flux-cut corrections.} 
It shows a clear trend that improving the angular resolution reduces positional uncertainties, thereby increasing the contrast between AGN–neutrino associations and random coincidences. At a fixed angular resolution, increasing the total event count enhances the statistical weight of the likelihood, roughly following the expected $ {(S/N)}_f \propto \sqrt{N}$ scaling. On the other hand, for a fixed number of detected neutrinos, the significance decreases with increasing PSF width by $ {(S/N)}_f \propto \sigma_{\rm psf}^{-1}$.
{Notably, reaching a $5\sigma$ detection requires substantially fewer events for an angular resolution of $\sim0.1^\circ$. By contrast, for angular resolution worse than $\sim0.4^\circ$, orders of magnitude more events are needed, making it difficult to identify the eFEDS-like population correlation.} 
{ The red star in Figure~\ref{fig:significance} indicates that the current IceCube alert-track sample is not expected to establish a statistically significant spatial correlation with a dense, eFEDS-like AGN population. Indeed, above 100~TeV, the present IceCube alert-track sample contains only $\sim 10^2$--$10^3$ astrophysical neutrinos and has a median angular uncertainty of $\sim 0.55^\circ$ \citep{IceCat-2}. Under these conditions, the cumulative signal from the broader eFEDS-like AGN population cannot be unambiguously separated from the isotropic background. This does not contradict the reported few-sigma stacking excesses from nearby X-ray-bright Seyferts \citep{Abbasi2022PhRvD, IceCube_ESH_Seyfert_South, IceCube_HardXray_AGN, IceCube_Xray_Bright_AGN}, because those searches mainly probe the bright end of the AGN population+. We discuss this complementary bright-source regime further in Sec.~\ref{sec:bright_sample}.} 

Figure~\ref{fig:significance} demonstrates the importance of angular resolution. 
First, a smaller $\sigma_{\rm psf}$ mitigates source confusion and allows the likelihood  analysis to incorporate contributions from a larger fraction of the AGN population, including sources with weaker X-ray emission, provided that the mean source separation satisfies $\bar{d} \geq 3\sigma_{\rm psf}$. In this regime, reducing $\sigma_{\rm psf}$ allows more sources to satisfy $\bar{d} \geq 3\sigma_{\rm psf}$, so that they can be included in the likelihood with non-negligible weights (the $w_j$ of Eq.~\ref{eq:S}), thereby enhancing the cumulative statistical power.
Second, improved angular resolution sharpens the spatial probability density of each neutrino event, concentrating the signal likelihood around the true source positions. This reduces the overlap with the isotropic background term and increases the contrast between genuine AGN–neutrino associations and random coincidences (i.e., the $S_i$ and $B_i$ of Eq.~\ref{eq:L_f}), leading to a larger likelihood ratio and a higher detection significance.

\begin{figure*}
    \centering
    \includegraphics[width=0.45\linewidth]{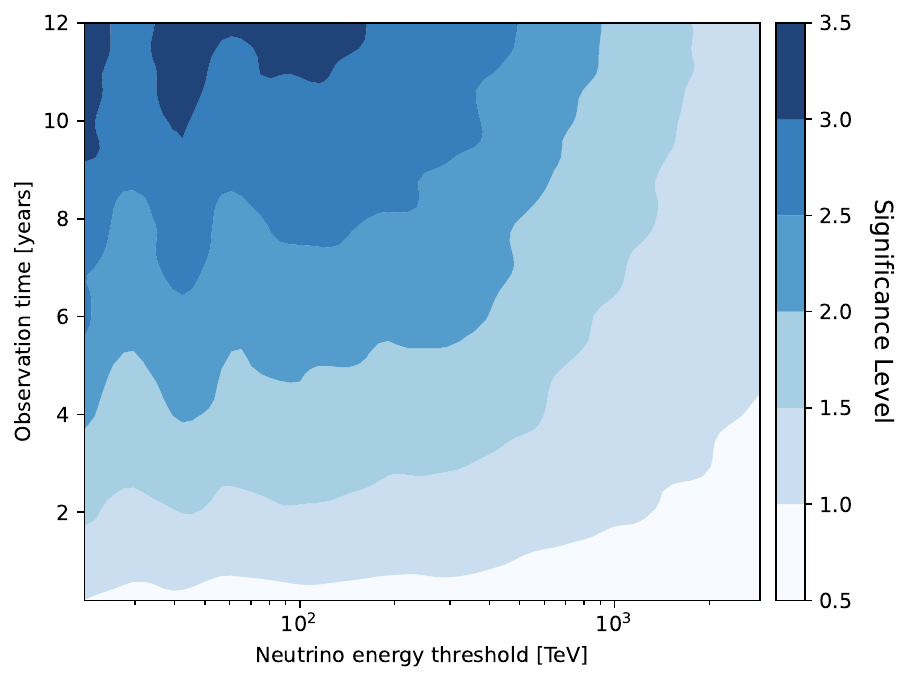}
    \includegraphics[width=0.45\linewidth]{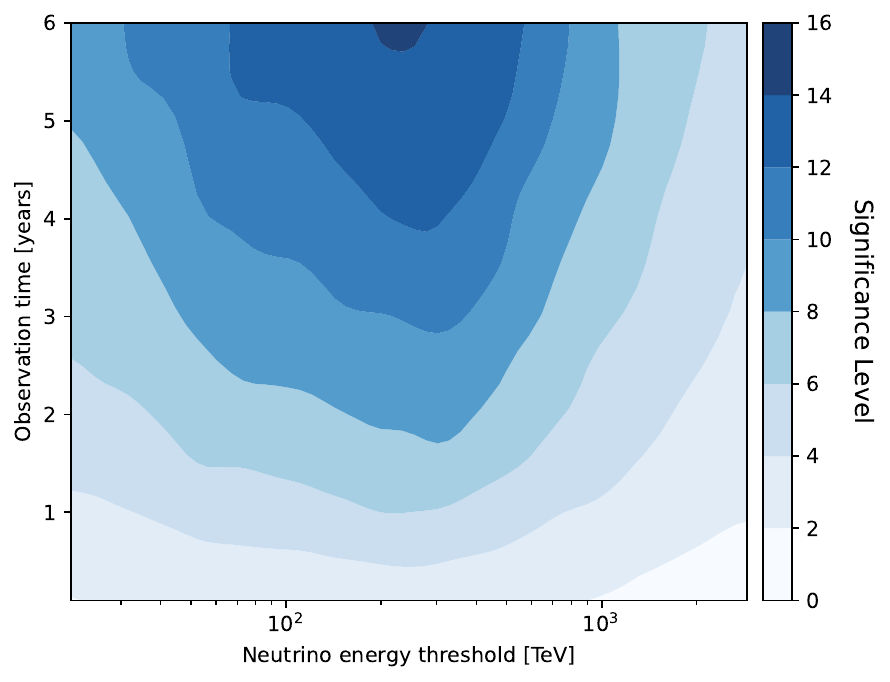}
    \caption{
    Trade-off between neutrino energy threshold and exposure time for future neutrino observatories.
    \textbf{The left panel} shows results for IceCube-Gen2, adopting the projected energy-dependent angular resolution from \citet{gen2ExPSF2019EPJWC.20705002B}. \textbf{The right panel} shows the corresponding results for the proposed 30\,km$^3$ HUNT detector, incorporating its energy-dependent angular resolution and effective area from \cite{HTQ2025icrc}.   {The color scale gives the expected detection significance after accounting for the eFEDS resolved fraction and the unresolved AGN contribution.}
   }
    \label{fig:ICE2_HUNT}
\end{figure*}

Based on the influence of angular resolution described above, we next examine how the selection of different neutrino energy bands influences the source identification capabilities. In particular, the fraction of detected neutrino events above a given energy threshold plays a key role in the achievable detection significance: astrophysical neutrinos from AGN are expected to dominate at higher energies, while the atmospheric background falls steeply. Moreover, higher-energy neutrinos are reconstructed with better angular resolution, further enhancing source identification capabilities.

Figure~\ref{fig:ICE2_HUNT} illustrates the trade-off between neutrino energy threshold and exposure time for future neutrino observatories, with the left panel corresponding to IceCube-Gen2 and the right panel to the proposed HUNT detector. 
The horizontal axis denotes the minimum reconstructed neutrino energy threshold applied to the event sample, while the vertical axis indicates the observation time. 
For IceCube-Gen2, we adopt an energy-dependent angular resolution based on muon track reconstruction reported in \citet{gen2ExPSF2019EPJWC.20705002B}, together with the corresponding effective area taken from the official IceCube-Gen2 design study \citep{IceCube-gen22021}. 
For the HUNT detector, the calculation incorporates its energy-dependent angular resolution and effective area from the simulations \citep{HTQ2025icrc}. {This corresponds to about $\sim1100$ events above 100~TeV and $\sim270$ events above 300~TeV per year for HUNT.}

As the energy threshold of neutrinos increases, two competing effects emerge: 
(i) the angular resolution improves and the atmospheric neutrino flux drops; 
(ii) the number of neutrino events above the energy threshold declines, reducing statistical power. These opposing trends jointly define an optimal energy window, typically around a few hundred TeV, where the sensitivity peaks
(100~TeV for IceCube-Gen2 and 300~TeV for HUNT).
Beyond this regime, further increases in the energy threshold result in diminishing returns due to limited event statistics.

Within the same physical framework, the superior performance of the HUNT detector can be understood in terms of its expected design characteristics. 
Based on its projected, energy-dependent performance, HUNT is anticipated to achieve a median angular resolution that is improved by a factor of $\sim$4–5 relative to IceCube-Gen2, together with an effective area larger by a factor of $\sim$3 in the relevant energy range (a few hundred TeV). 
These factors summarize the relative detector performance inferred from their respective design specifications, rather than representing a direct rescaling of IceCube-Gen2 parameters. 
As a result, HUNT attains a higher detection significance—or equivalently, requires a shorter exposure time—for a given neutrino energy threshold.

\section{Impact of Systematic and Positional Uncertainties} \label{sec:uncer}

The analyses presented thus far have assumed idealized conditions: the AGN X-ray luminosity function was treated as perfectly known, and the neutrino angular uncertainty was assumed to be fixed for all events. In practice, both carry uncertainties, which propagate into the inferred total AGN fraction $f_{\rm tot}$. Here we quantify the contributions from these two dominant sources of uncertainty and evaluate their combined impact on the overall detection significance. In addition to these systematic effects, statistical fluctuations due to the finite neutrino sample are automatically incorporated in all Monte Carlo realizations, and are not treated as a separate source of uncertainty. As the exposure increases, the impact of these statistical fluctuations diminishes, and the total uncertainty becomes increasingly dominated by systematic effects.

The first source of systematic uncertainty arises from extrapolating the neutrino contribution inferred from the flux-limited AGN subset ($F_X > F_{\rm min}$) to the full AGN population.  {The likelihood fit uses the observed eFEDS flux distribution to define the resolved template, while the XLF is used to correct for AGNs not represented by this template, including fainter unresolved sources and the rare bright sources outside the small eFEDS field.}

\begin{figure}
    \centering
    \includegraphics[width=0.48\textwidth]{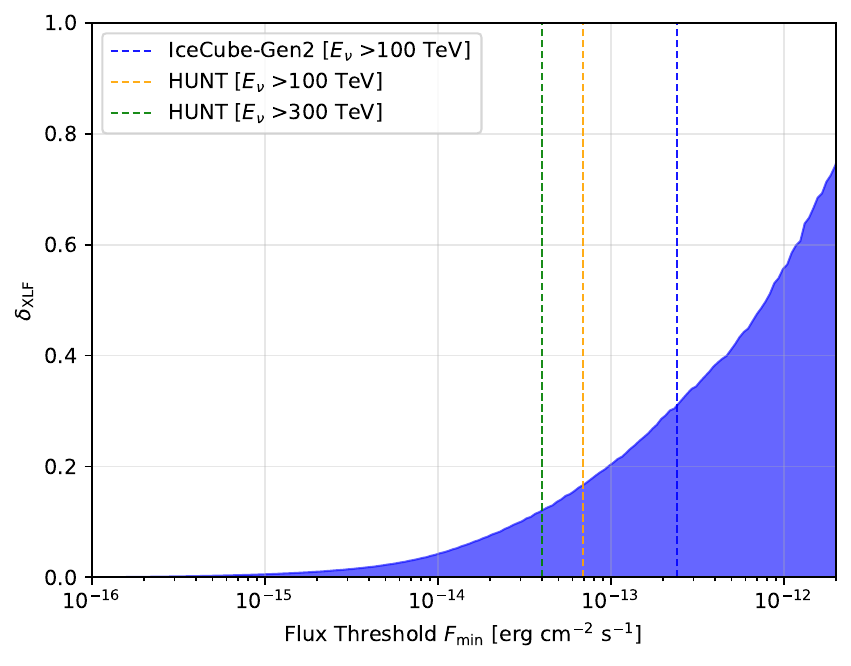}
    \caption{
    Relative uncertainty factor $\delta_{\rm XLF}$ in the extrapolated AGN contribution as a function of the AGN flux threshold $F_{\rm min}$. The shaded band shows the 68\% confidence interval derived from Monte Carlo realizations. {The vertical blue, orange, and green lines mark the adopted $F_{\rm cut}$ values for IceCube-Gen2 ($E_\nu>100$~TeV), HUNT ($E_\nu>100$~TeV), and HUNT ($E_\nu>300$~TeV), respectively.}  The uncertainty remains modest at low thresholds but grows rapidly when only the brightest AGNs are included. } 
    \label{fig:XLF_uncertainty}
\end{figure}

Here we quantify the systematic uncertainty associated with this extrapolation. Assuming that the neutrino luminosity scales linearly with the X-ray luminosity \citep[e.g.,][]{Abbasi2022PhRvD, Kun2024PhRvD}, the total neutrino flux can be written as
\begin{equation}
\Phi_\nu^{\rm tot} \propto \int_{L_{\rm min}}^{L_{\rm max}} L_X \, \phi(L_X,z)\, dL_X\,dz ,
\end{equation}
where $\phi(L_X,z)$ denotes the XLF.
Using the best-fit XLF parameters and their reported $1\sigma$ uncertainties from \citet{UedaXLF2014}, we perform Monte Carlo simulations in which the XLF parameters are randomly varied according to independent Gaussian distributions centered on their best-fit values. For each realization, we compute the corresponding correction factor between the flux-limited subset and the full population, thereby estimating the relative uncertainty of the extrapolated neutrino contribution as a function of the adopted flux threshold $F_{\rm min}$. {In the practical application to certain detector and AGN catalog, this threshold is the flux cut $F_{\rm cut}$ set by the source-separation criterion.} The uncertainties in the XLF is then involved in the extrapolated total AGN neutrino, which we denote as $\delta_{\rm XLF}$.
The resulting $\delta_{\rm XLF}$ as a function of $F_{\rm min}$ is summarized in Figure~\ref{fig:XLF_uncertainty}. As expected, adopting brighter flux thresholds leads to rapidly increasing extrapolation uncertainties. This behavior reflects the fact that when only very luminous AGNs are included, the selected subset captures only a small fraction of the total XLF-weighted luminosity density, causing the correction factor to become highly sensitive to uncertainties in the faint-end slope and normalization of the XLF.

As discussed in Sec.~\ref{sec:llh_analysis} and Sec.~\ref{sec:mitigate}, the total AGN neutrino fraction $f_{\rm tot}$ is obtained by applying the flux-threshold and XLF extrapolations to the flux-limited likelihood result, as defined in Eq.~(\ref{eq:f_total}). 
Uncertainties in the XLF-based extrapolation propagate directly into the inferred total AGN neutrino fraction $f_{\rm tot}$. 
In the Monte Carlo framework, this appears as a broadening of the reconstructed $f_{\rm tot}$ distribution, characterized by an increased standard deviation $\sigma_{f_{\rm tot}}$, while the mean remains approximately unbiased. 
We quantify the impact of this effect using the relative uncertainty $\delta$. 
When $\delta > 50\%$, the inference of $f_{\rm tot}$ becomes dominated by XLF systematic uncertainty rather than statistical fluctuations, and the extrapolated AGN contribution is no longer robustly constrained.
To avoid this regime, we restrict our analysis to flux thresholds 
$F_{\rm cut} < 7\times10^{-13}\,\mathrm{erg\,cm^{-2}\,s^{-1}}$, 
which corresponds to the stability criterion quantified above, as shown in Figure~\ref{fig:XLF_uncertainty}. 
In this range, the selected AGN subset encompasses a substantial fraction of the total XLF-weighted luminosity density, and the extrapolation remains stable against variations in the XLF parameters.

The second uncertainty arises from the event-by-event variations in the $\sigma_{\rm psf}$. In previous sections, we adopted a fixed PSF width $\sigma_{\rm psf}$ for given energy bins, whereas real detectors exhibit a distribution of reconstructed angular uncertainties. This variation effectively broadens the signal and background probability density functions used in the maximum-likelihood analysis, thereby reducing sensitivity. To quantify this effect, we simulate neutrino samples by assigning to each event a $\sigma_{\rm psf}$ randomly sampled from the detector’s energy-dependent angular resolution distribution. We find that the uncertainty induced by event-by-event PSF variations is smaller than that from the XLF-based extrapolation, and therefore does not dominate the total error budget.

\begin{figure}
    \centering
    \includegraphics[width=0.48\textwidth]{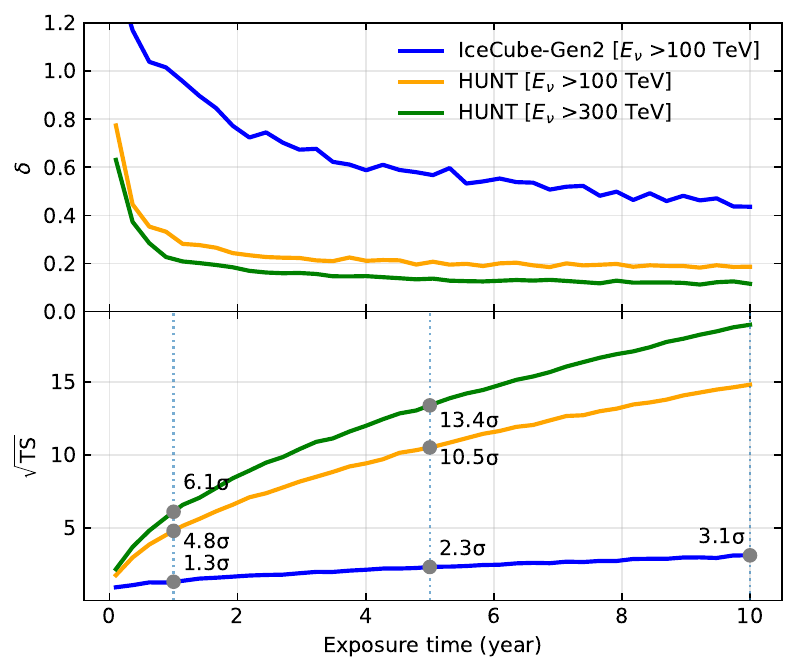}
    \caption{
    \textbf{Top panel:} Relative uncertainty on the inferred total AGN neutrino fraction $\delta$ as a function of the exposure time, derived from repeated Monte Carlo simulations.
    \textbf{Bottom panel:} The  detection significance, expressed as $\sqrt{\rm TS}$, as a function of exposure time. The steady decrease of $\delta$ and the growth of $\sqrt{\rm TS}$ demonstrate how both the precision and statistical significance of the AGN contribution improve with increased exposure.}
    \label{fig:ExposureTime_sigma}
\end{figure}

Combining the uncertainties from both XLF extrapolation and PSF variation, we assess the sensitivity of future neutrino telescopes to testing the AGN-origin hypothesis. In the simulations, the angular uncertainty of each neutrino event is treated on an event-by-event basis, with the reconstructed direction randomly drawn according to the corresponding point-spread function, thereby accounting for realistic variations in the localization accuracy.
Focusing on events with energies above 100\,TeV  ($\beta \simeq 0.5$), IceCube-Gen2 is expected to achieve a median angular resolution of $\sim0.2^\circ$. 
To avoid source confusion, we therefore require the selected AGN population to satisfy the condition $\bar{d} \geq 0.6^\circ$.
Based on the eFEDS AGN catalog over the $142\,\mathrm{deg}^2$ survey area, this requirement translates into a maximum of $\sim150$ AGNs within the field of view. 
This criterion selects the brightest sources with X-ray fluxes {$F_X>F_{\rm cut}=2.4\times10^{-13}\,\mathrm{erg\,s^{-1}\,cm^{-2}}$,} which are sufficiently sparse to remain well resolved at the IceCube-Gen2 angular resolution. 
As shown in Figure~\ref{fig:XLF_uncertainty}, for the {$F_{\rm cut}$} value corresponding to IceCube-Gen2 (blue vertical line), the uncertainty induced by the XLF-based extrapolation from the flux-limited AGN template to the full AGN population is $\delta_{\rm XLF} \simeq 30\%$.
{As shown in Figure~\ref{fig:ExposureTime_sigma}, such a detector would achieve a significance of approximately $\sim2.3\sigma$ after five years and $\sim3.1\sigma$ after ten years of exposure in testing whether all astrophysical neutrinos originate from the AGN population. This significance is suggestive but remains below the level of a decisive discovery.}

As an illustrative and more optimistic case, we consider the HUNT, a proposed 30\,km$^3$ underwater neutrino detector \citep{HUNTproposal2023, HTQ2025icrc} and focus on events with energies above 300\,TeV ($\beta \simeq 0.8$), for which an angular resolution of $\sim0.04^\circ$ is expected at these energies. 
At this energy threshold, the source-separation criterion selects AGNs with X-ray fluxes {$F_X>F_{\rm cut}=4\times10^{-14}\,\mathrm{erg\,s^{-1}\,cm^{-2}}$.} 
Based on the eFEDS AGN catalog, this $\bar{d} \geq 3\sigma_{\rm psf}$ condition corresponds to a maximum surface density of $\sim 3200$ sources.
Such a relatively low flux threshold significantly reduces the uncertainty associated with the XLF-based extrapolation. As indicated by the green vertical line in Figure~\ref{fig:XLF_uncertainty}, the corresponding relative uncertainty is $\delta_{\rm XLF}\simeq 10\%$.  
{We find that HUNT can test the AGN-origin hypothesis at a confidence level of about $6.1\sigma$ after one year of exposure, after accounting for the eFEDS resolved fraction, XLF extrapolation, and PSF variation, as shown in Figure~\ref{fig:ExposureTime_sigma}.
At the same time, the likelihood analysis constrains the total AGN neutrino fraction to a relative precision of $\delta \simeq 20\%$. With five years of exposure, the significance further increases to $\sim 13.4\sigma$, and the relative uncertainty decreases to $\delta \simeq 13\%$.}

For comparison, we also consider a more conservative energy threshold of $E_\nu > 100$\,TeV for HUNT. In this case, the larger angular uncertainty requires a brighter AGN flux threshold {$F_{\rm cut}$} (orange vertical line in Figure~\ref{fig:XLF_uncertainty}), leading to a larger XLF extrapolation uncertainty and consequently a lower detection significance (orange curve in Figure~\ref{fig:ExposureTime_sigma}), although the AGN-origin hypothesis remains testable at high confidence with multi-year exposure. {For this threshold, the expected significance is $\sim4.8\sigma$ after one year and $\sim10.5\sigma$ after five years.}

\section{Discussion}

\subsection{AGNs contributing partially to the total astrophysical neutrino flux}

\begin{figure}
\centering
    \includegraphics[width=0.45\textwidth]{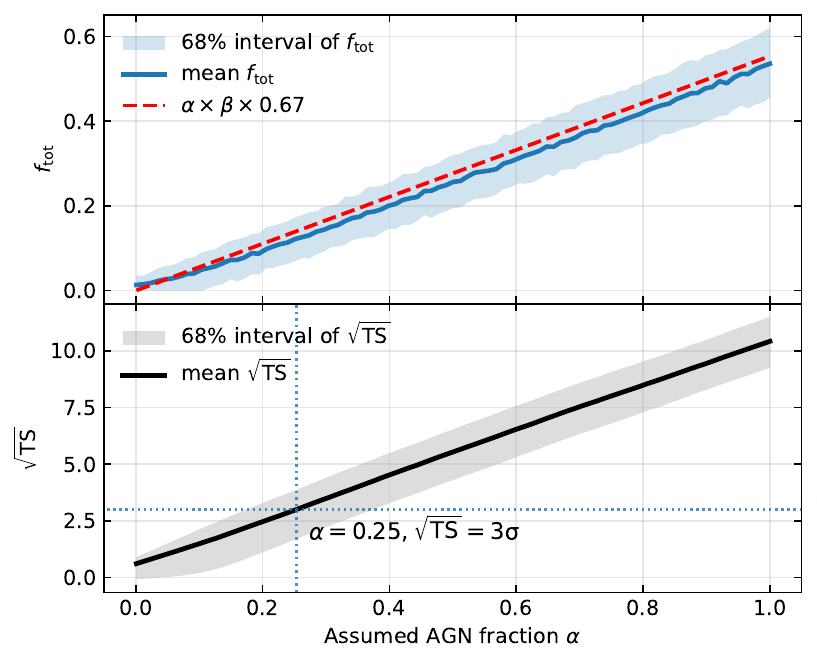}
    \caption{ 
    \textbf{Top panel:} Recovered eFEDS-resolved AGN template fraction as a function of the assumed AGN neutrino fraction $\alpha$. The blue shaded band shows the central 68\% interval from 800 Monte Carlo realizations, while the solid blue line indicates the mean recovered fraction from the simulations. {The red dashed line represents the expected resolved-template contribution, $0.67\,\alpha\beta$, after accounting for the eFEDS flux completeness.}
    \textbf{Bottom panel:} Corresponding detection significance $\sqrt{\rm TS}$ as a function of $\alpha$, showing that even partial AGN contributions can be detected when the resolved template component is sufficiently large.}
    \label{fig:disc1}
\end{figure}

Throughout most of this work we have assumed the limiting case in which AGNs are assumed to be the dominant contributors to the observed astrophysical neutrino flux. In reality, the diffuse neutrinos could be produced by a mixture of source populations, and AGNs may account for only a fraction of the total astrophysical neutrino flux. {Our framework can also treat the more general case in which AGNs produce only a fraction of the astrophysical neutrino flux. This fraction is described by $\alpha$. In the spatial-correlation test, only the eFEDS-resolved part of this AGN component is assigned to cataloged sources. The unresolved AGN component and the non-AGN neutrinos are both treated as diffuse background.}

Figure~\ref{fig:disc1} demonstrates how the detection significance $\sqrt{\rm TS}$ scales with the assumed AGN contribution $\alpha$ for a representative next-generation detector scenario (a 30\,km$^3$-scale underwater neutrino telescope with 3 years of exposure). 
{The top panel shows the recovered eFEDS-resolved template contribution as a function of $\alpha$, obtained from repeated Monte Carlo simulations. Because the eFEDS-depth template resolves only about $67\%$ of the total AGN flux, the recovered spatially correlated component follows the expectation $0.67\,\alpha\beta$. The remaining unresolved AGN contribution is absorbed into the diffuse component.}
The bottom panel displays the corresponding detection significance, expressed as $\sqrt{\rm TS}$, as a function of $\alpha$. { The vertical marker indicates the value $\alpha \simeq 0.25$, where $\sqrt{\rm TS} = 3$. This means that if AGNs contribute less than $\sim 25\%$ of the total astrophysical neutrino flux, the 3-year HUNT exposure would be insufficient to achieve a 3$\sigma$ detection.}
For larger AGN contributions, $\sqrt{\rm TS}$ scales approximately linearly with the resolved template contribution, reflecting the fact that the likelihood analysis primarily probes the spatially correlated AGN component, while the unresolved component behaves as part of the diffuse background.

\subsection{X-ray flux-weighting}

\begin{figure}
\centering
    \includegraphics[width=0.45\textwidth]{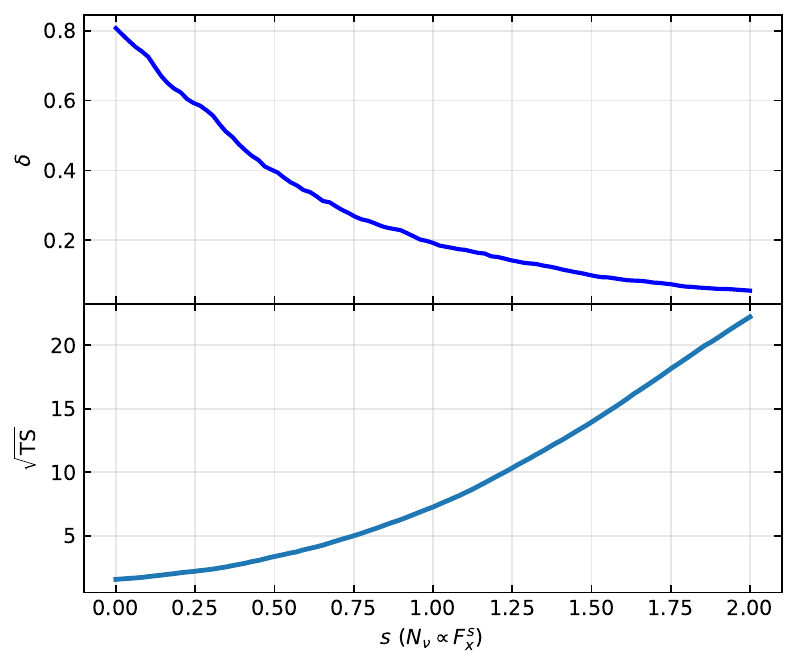}
    \caption{ 
    \textbf{Top panel:} Relative uncertainty of the recovered AGN neutrino fraction $\delta$ as a function of the flux-weighting parameter $s$ ($N_\nu \propto F_X^s$).
    \textbf{Bottom panel:} Detection significance $\sqrt{\rm TS}$ as a function of the flux-weighting parameter $s$, illustrating that stronger weighting toward luminous AGNs enhances the detectability.  
    All panels correspond to a 30\,km$^3$ undersea neutrino telescope with 3 years of exposure.}
    \label{fig:disc2}
\end{figure}

So far, we have assumed a linear relationship between the number of neutrinos and the AGN X-ray flux. This treatment considers that the X-ray flux of an AGN may be a proxy for its high-energy proton power and also representative for the target density for neutrino production. However, the linear relationship is unwarranted yet \citep[e.g.,][]{Berezinsky2024,Saurenhaus2025arXiv250706110S}.
To generalize this idea, we explore a family of models in which the neutrino flux from an AGN scales with its X-ray flux as $N_\nu \propto F_X^s$. The parameter $s$ reflects the assumed physical correlation strength: $s = 0$ corresponds to no dependence (i.e., uniform weighting across AGNs), while increasing $s$ emphasizes the role of brighter AGNs in neutrino production.
In this case, the source weights in Equation~\ref{eq:S} of the likelihood function are modified to  $w_j = {F_j^s}/{\sum_j F_j^s}$ ensuring that the simulated neutrino--AGN associations reflect the chosen correlation index $s$. {For consistency, the flux-threshold correction is also generalized by replacing $F_X$ with $F_X^s$ in Eq.~(\ref{eq:f_total}).}

We also consider a special scenario in which all astrophysical neutrinos originate from AGNs, corresponding to $\alpha = 1$.
Figure~\ref{fig:disc2} summarizes the impact of the flux-weighting parameter $s$ on both the precision of the recovered AGN neutrino fraction and the overall detection significance, for a 30\,km$^3$ neutrino telescope with three years of exposure.
The top panel of Figure~\ref{fig:disc2} displays the relative uncertainty of the recovered AGN neutrino fraction $\delta$. We find that $\delta$ decreases monotonically with increasing $s$, indicating that the precision of the recovered AGN contribution improves as neutrino emission becomes more strongly associated with X-ray--bright sources. 
The bottom panel shows the detection significance, quantified by $\sqrt{\rm TS}$, as a function of $s$. A clear monotonic increase is observed, indicating that stronger weighting toward luminous AGNs substantially enhances the detectability of the population-level correlation. In the limit $s \simeq 0$, corresponding to uniform source weighting, the detection significance remains modest, at the level of $\sim1$--$2\sigma$. {For the baseline linear case $s=1$, the significance is about $7\sigma$ in this three-year HUNT-like setup. The significance becomes even higher if the neutrino output is more strongly concentrated in X-ray-bright AGNs.}
This behavior can be understood as a consequence of two related effects. First, larger values of $s$ effectively suppress the contribution from faint AGNs while simultaneously increasing the relative weight of X-ray--bright sources, which significantly enhances the detection significance and reduces the statistical uncertainty in the recovered AGN neutrino fraction. Second, smaller values of $s$ assign a relatively larger weight to faint AGNs, for which the extrapolation of the XLF introduces substantial systematic uncertainties, further inflating the uncertainty in the recovered fraction.

Overall, these results suggest that if astrophysical neutrino production preferentially traces the most luminous AGNs, both the detectability of the AGN--neutrino correlation and the precision of the inferred AGN neutrino fraction are significantly improved. Conversely, weaker flux weighting ($s < 0.5$) implies a larger relative contribution from faint AGNs, leading to reduced signal contrast with respect to the isotropic background and increased uncertainty in the recovered fraction.

\subsection{Blazars Neutrino Predictions}

\begin{figure*}
\centering
    \includegraphics[width=0.45\textwidth]{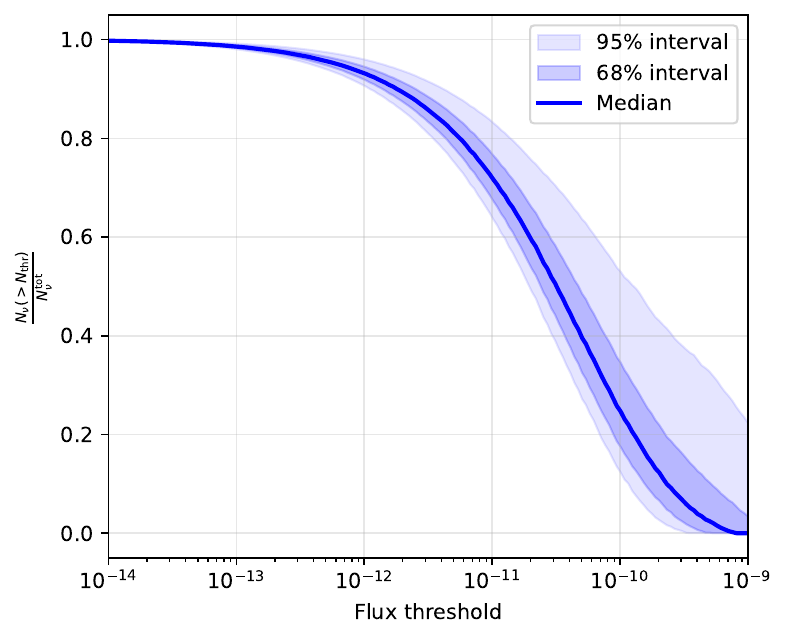}
    \includegraphics[width=0.44\textwidth]{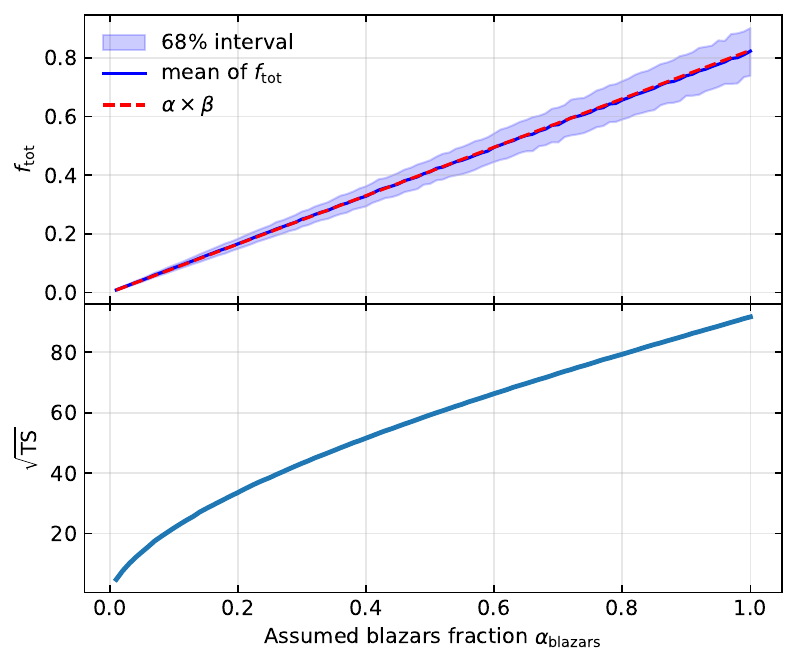}
    \caption{ 
    \textbf{Left:} Cumulative fraction of the total neutrinos contributed by blazars as a function of the gamma-ray flux  threshold of blazars. The solid line shows the median, while the shaded regions indicate the 68\% and 95 \% credible intervals.
    \textbf{Right:} Top panel: Same as the Figure~\ref{fig:disc1}, recovered blazar neutrinos fraction $f_{\rm tot}$ as a function of the blazar neutrino fraction $\alpha_{\rm blazars}$. 
    Bottom panel: Corresponding detection significance $\sqrt{\rm TS}$ as a function of $\alpha_{\rm blazars}$, showing that even partial blazars contributions can be detected with high statistical significance.}
    \label{fig:blazars}
\end{figure*}

Blazars are also widely discussed as possible sources of extragalactic neutrinos, and are likely to contribute a significant fraction in the diffuse neutrino intensity at sub-PeV energies  (e.g., \citealt{Murase2014blazar,2018SciTXS0506,Saikat2022, Buson2022blazar}).
Motivated by the association of TXS~0506+056 with a hard-spectrum ($\Gamma \approx 2$) neutrino flare, we focus on the contribution of blazars to neutrinos above $\sim100$~TeV, where such sources are expected to be more prominent.

To directly test this scenario, IceCube has performed dedicated searches for cumulative emission from GeV-selected blazar populations \citep{ICblazar2017ApJ}. No significant excess was observed, and the upper limits constrain the contribution of 2LAC blazars to $\lesssim27\%$ of the observed diffuse neutrino flux between 10~TeV and 2~PeV (for a spectral index $\Gamma=2.5$). Nevertheless, it is reasonable to expect that their relative contribution becomes more significant above $\sim100$~TeV, and hence we focus on this higher-energy regime to assess the role of blazars as neutrino sources.

We consider  3933  blazars from the fourth Fermi-LAT catalog \citep{4FGLDR4},  which provide a representative sample for our study.
We also extrapolate their collective neutrino contribution using the GeV gamma-ray luminosity function (GLF) \citep{BlazarsGLF2020PhRvD}, following the same procedure as that applied to the AGN XLF in Section~\ref{sec:uncer}.
The left panel of Figure~\ref{fig:blazars} presents the cumulative fraction of the total blazar neutrino flux as a function of flux threshold, with shaded regions showing the 68\% and 95\% credible intervals.
Here, the ``flux threshold''  corresponds to the average energy flux detection limit of Fermi-LAT, representing the approximate completeness threshold of the blazar sample. Given that the mean sensitivity limit for the Fermi-LAT blazar sample is about $2\times10^{-12}\,\mathrm{erg\,s^{-1}\,cm^{-2}}$ \citep{fermicat2022ApJS}, we estimate that this Fermi-LAT blazar population accounts for approximately $\sim85\%\pm5\%$ of the total neutrino emission from all blazars. This result is broadly consistent with previous Fermi-LAT studies of the blazar population at high energies. In particular, analyses of the extragalactic $\gamma$-ray background above $\sim 50$~GeV have shown that the bulk of the total $\gamma$-ray energy density is dominated by resolved blazars, while the contribution from unresolved, low-flux sources is subdominant \citep[e.g.,][]{FermiEGB50GeV2016PRL}. Similar conclusions have been reached through stacking analyses of faint and extreme blazar populations, which indicate that unresolved sources contribute only a modest fraction of the total $\gamma$-ray flux \citep{Fermibalzarstack2019}.

The right panel of Figure~\ref{fig:blazars}  illustrates the trend of the detection significance as a function of the assumed blazars fraction for the HUNT, a 30\,km$^3$ detector over 3 years, with systematic uncertainties included. 
When systematic effects are included, the dominant limitation arises from GLF uncertainties. In this case, if blazars contribute to $\sim$5\% of the diffuse neutrino flux above 300~TeV, our method reaches a $\sqrt{\rm TS}>9$ detection significance for 3 years of exposure. For larger assumed fractions, the significance saturates, since the GLF uncertainties impose a systematic floor on the achievable sensitivity. This demonstrates that our approach is capable of identifying even a modest blazar contribution to the diffuse neutrino flux with high confidence.

{
\subsection{Consistency Check with Current IceCube Seyfert Searches}\label{sec:bright_sample}

Recent IceCube analyses have explored the possible connection between some AGNs and high-energy neutrinos. 
Several stacking and collective searches of nearby Seyfert galaxies selected from the Swift/BAT catalog have reported excesses at the $\sim 3\sigma$ level  \citep{Abbasi2022PhRvD, IceCube_ESH_Seyfert_South, IceCube_HardXray_AGN, IceCube_Xray_Bright_AGN}. 
These studies are primarily sensitive to the brightest AGNs, such as NGC~1068 and NGC~4151, and suggest that some X-ray–bright AGNs may act as neutrino emitters. 
These results show that current IceCube observations are beginning to probe the bright end of the AGN population.

We have verified our simulation framework by applying a stacking analysis to northern-sky Swift/BAT Seyfert AGNs, ranked and weighted by the intrinsic 2--10~keV X-ray fluxes from the BAT AGN Spectroscopic Survey \citep[BASS;][]{BATagn2017ApJS}. 
We reproduce a comparable level of statistical significance (see Appendix~\ref{app:bat_stacking} for details).
This comparison provides a natural interpretation of current observations. In our simulations, the total contribution of the AGN population to the astrophysical neutrino flux is a free parameter. The brightest nearby sources contribute only a minor fraction of the total AGN neutrino output \citep{2025arXiv251013403A,ic2025ApJ...988..141A}, while the rest majority is distributed among many fainter AGNs. As a result, current IceCube analyses can be sensitive to the bright-source component, while the cumulative signal from the much larger faint AGN population remains difficult to detect with high significance. Our results show that future detectors can identify this broader population component through spatial correlations with high-energy neutrinos.
}

\section{Conclusions}

We have evaluated the prospect of diagnosing the correlation between X-ray AGNs and high-energy neutrinos  with future neutrino telescopes, incorporating both statistical and systematic effects. By combining a flux-weighted likelihood method with realistic detector performance and observational constraints, we have found that:
\begin{itemize}
    \item \textbf{Dominant role of angular resolution:}  
    The sensitivity of AGN--neutrino correlation  is primarily driven by the  point-spread function of neutrino detectors. Achieving sub-degree resolution, particularly $\lesssim 0.1^\circ$, dramatically reduces background contamination and enhances the significance of detection. {For example, with sub-$0.1^\circ$ angular resolution, a sample of several thousand neutrinos are enough to reach a $5\sigma$ detection of the AGN--neutrino spatial correlation. In contrast, for angular resolutions worse than $\sim0.4^\circ$, random coincidences strongly dilute the signal, and orders of magnitude more events would be required.}

    \item \textbf{Source separation criterion:} 
    To minimize source confusion and ensure robust association between neutrinos and AGNs, we require the mean angular separation between two selected AGNs to exceed three times of the detector's PSF width (i.e., $\bar{d} > 3\sigma_{\rm psf}$). This criterion effectively avoids source confusion.

    \item \textbf{Energy threshold optimization:}  
    Increasing the minimum neutrino energy threshold improves the angular resolution and suppresses the atmospheric neutrino background, but at the cost of reducing the event statistics. For a 30\,km$^3$ undersea/underwater neutrino telescope (e.g., HUNT) that features rapidly improving angular resolution with energy (down to $\sim0.04^\circ$ above 300~TeV), the optimal sensitivity is reached at a threshold of a few hundred TeV. In contrast, IceCube-Gen2 shows a weaker energy dependence in its angular resolution; for this detector, selecting $E_\nu > 100$~TeV strikes a better balance between the angular precision and event statistics.

    \item \textbf{Flux-limited AGN extrapolation and XLF systematics:}  
    Our likelihood analysis is performed on a flux-limited AGN subset selected to satisfy $\bar{d} > 3\sigma_{\rm psf}$, ensuring that source confusion does not bias the fitted template fraction $f$. The total neutrino contribution from the full AGN population, $f_{\rm tot}$, is then obtained by applying the flux-threshold and XLF corrections. The associated uncertainty is strongly controlled by the adopted flux threshold $F_{\rm min}$, { or $F_{\rm cut}$ when applying the source-confusion criterion to practical detector.} For $F_{\rm min} \leq 10^{-14}\,{\rm erg\ cm^{-2}\ s^{-1}}$, the extrapolation uncertainty remains small ($\leq 10\%$), whereas for $F_{\rm min} \geq 7\times10^{-13}\,{\rm erg\ cm^{-2}\ s^{-1}}$ the inferred total contribution becomes unreliable, as the correction factor is dominated by uncertainties in the faint-end slope and normalization of the XLF.

    \item \textbf{Implications for next-generation neutrino telescopes:} 
    { Within the eFEDS-based population-correlation framework considered here,} a 30\,km$^3$-scale underwater neutrino telescope can reach a significance of {$6.1\sigma$ above 300\,TeV} with one year of exposure. With five years of exposure, the significance increases to {$\sim 13.4\sigma$}, and the total AGN neutrino fraction can be constrained to a relative precision of {$\delta \simeq 13\%$}. For IceCube-Gen2, using neutrinos above 100~TeV, the expected significance is {$\sim3.1\sigma$ with ten years of exposure}.

    \item \textbf{Sensitivity to partial AGN contributions:}  
        For a 30\,km$^3$-scale underwater neutrino telescope, an AGN contribution of {$\alpha \leq0.25$} to the astrophysical neutrino flux would not be distinguishable from an isotropic background at the $3\sigma$ level with three years of exposure. {For larger contributions, the detection significance scales approximately with the eFEDS-resolved template component.} 
\end{itemize}

These results demonstrate that next-generation neutrino observatories with sub-degree angular precision and sufficient exposure have the potential to confirm or strongly constrain the AGN origin of high-energy neutrinos. Deep multi-wavelength AGN surveys, especially in X-rays, are crucial to minimize uncertainties in the flux extrapolation and to fully exploit the discovery potential of upcoming facilities. 
{ More generally, the framework developed here is not tied to the eFEDS catalog or to a specific X-ray band. Future all-sky and wide-area X-ray surveys, including SRG/eROSITA \citep{eRASS1AGN2025} and Athena/WFI \citep{2023ExA_Athena}, will provide more complete AGN samples and make population-level AGN--neutrino correlation tests more powerful.}

\begin{acknowledgments}
We would like to thank Tian-Qi Huang and Shiqi Yu for valuable discussions. This work
is supported by the National Natural Science Foundation of China (grant Nos. 12333006, 12121003  and  12393852),  National SKA Program of China (2025SKA0110104) and the Fundamental Research Funds for the Central Universities (KG202502). We are grateful to the High Performance Computing Center of
Nanjing University for doing the numerical calculations in this
paper on its blade cluster system.
\end{acknowledgments}

\appendix

{
\section{Robustness to Source-to-source Scatter in the X-ray--Neutrino Relation}
\label{app:scatter}

The baseline analysis assumes that the AGN neutrino luminosity is traced by the X-ray flux at the population level. In reality, individual sources may deviate from this simple scaling because of intrinsic source-to-source variation in the X-ray--neutrino relation \citep{Jain2026}. We therefore perform an additional robustness test to quantify how such scatter affects the projected correlation sensitivity.

We model this source-to-source variation by multiplying the nominal X-ray-based neutrino weight of each AGN by a random log-normal factor $\eta_j$, defined by
\begin{equation}
\ln \eta_j \sim \mathcal{N}(0,\sigma_{\ln F}^{2}) .
\end{equation}
Here $\sigma_{\ln F}$ characterizes the intrinsic scatter in log-space. In each Monte Carlo realization, a new set of $\eta_j$ values is randomly drawn for all AGNs. 
Since the weights are normalized after drawing the random scatter, the total injected AGN neutrino fraction is kept fixed; the scatter only redistributes the relative neutrino contribution among individual AGNs.

In the likelihood analysis, however, we deliberately use the unscattered X-ray-flux template,
without passing the realization-dependent $\eta_j$ values to the fit. This procedure avoids using hidden simulation information and tests how much sensitivity is lost when the true neutrino weights differ from the X-ray weights assumed in the analysis.

We carry out this test for the HUNT-like benchmark configuration with three years of exposure, $E_\nu>300$~TeV, event-wise random PSF values, and $800$ Monte Carlo realizations for each value of $\sigma_{\ln F}$. We consider $\sigma_{\ln F}=0$, $0.5$, $1.0$, and $1.7$.

\begin{figure*}
\centering
\includegraphics[width=0.47\textwidth]{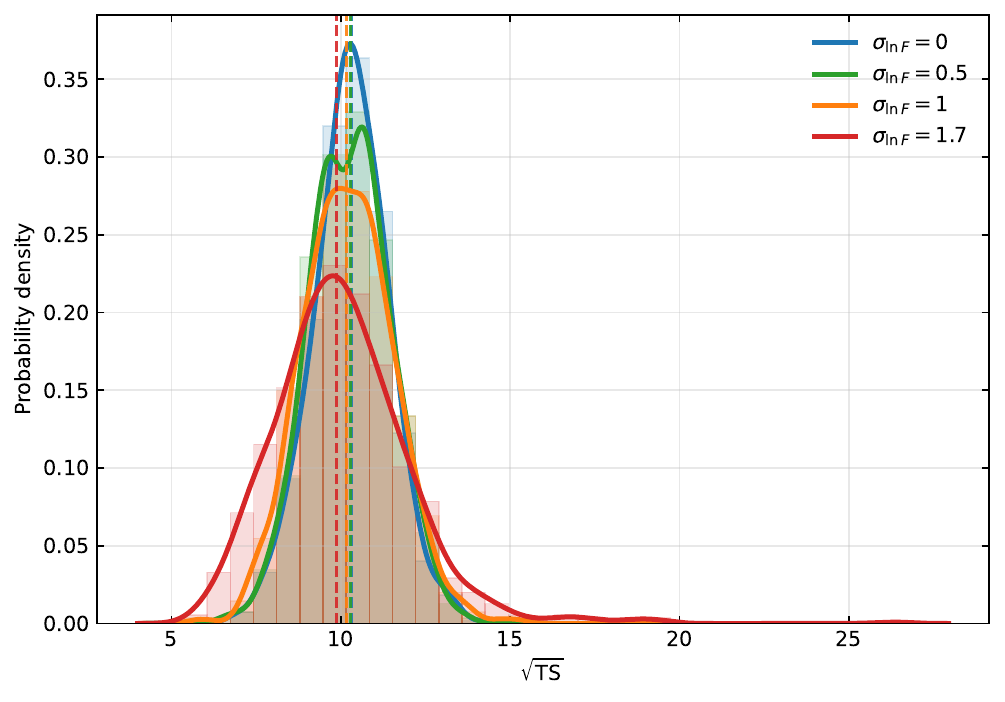}
\includegraphics[width=0.47\textwidth]{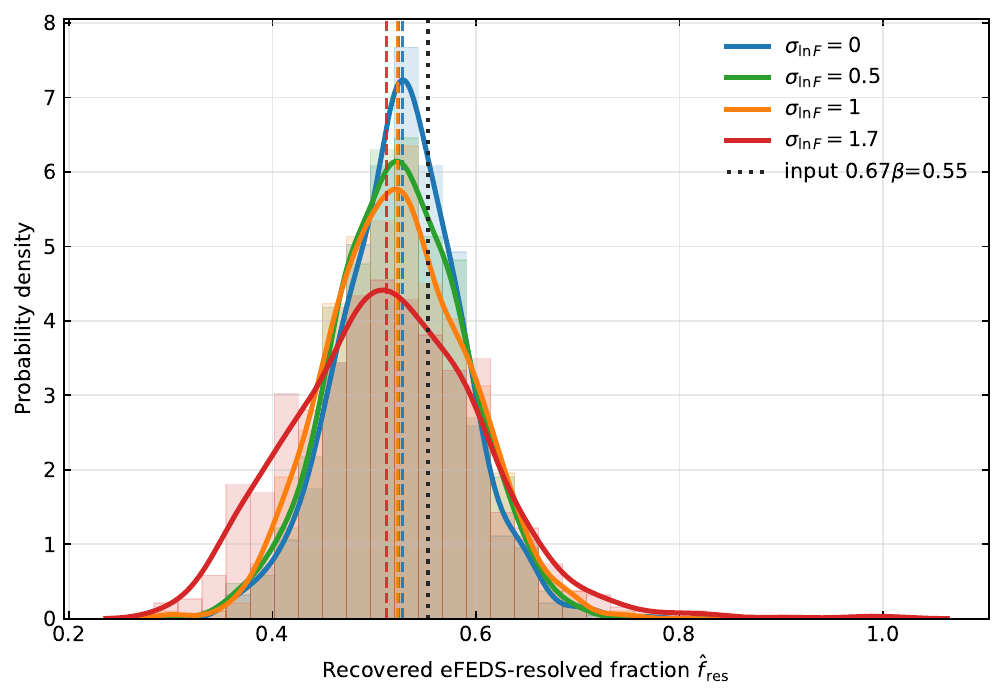}
\caption{
Distributions of the recovered detection significance and eFEDS-resolved template fraction in the source-to-source scatter test.
\textbf{Left:} distribution of $\sqrt{\rm TS}$ for different values of $\sigma_{\ln F}$.
\textbf{Right:} distribution of the recovered resolved-template fraction $\hat f_{\rm res}$ for the same realizations. The vertical dotted line marks the input value $0.67\beta=0.5$ for the HUNT $E_\nu>300$~TeV benchmark.
}
\label{fig:scatter_randompsf}
\end{figure*}

As shown in Figure~\ref{fig:scatter_randompsf}, moderate source-to-source scatter has little impact on the median detection significance. Even for the large-scatter case $\sigma_{\ln F}=1.7$, the median significance decreases only mildly, from $\sqrt{\rm TS}=10.32$ in the no-scatter case to $\sqrt{\rm TS}=9.86$. The main effect of the scatter is to broaden the realization-to-realization distribution, especially for the recovered resolved-template fraction.

We therefore interpret the baseline results as the optimistic case in which the observed X-ray flux provides a good population-level proxy for the neutrino flux. The scatter test shows that substantial source-to-source variation increases the uncertainty of the forecast, but does not qualitatively change the detectability of the eFEDS-resolved population-level AGN--neutrino spatial correlation for a HUNT-like detector.

}

\section{Background TS Distribution and Significance Calibration}\label{sec:ts_bg}

\begin{figure}
    \centering
    \includegraphics[width=0.5\textwidth]{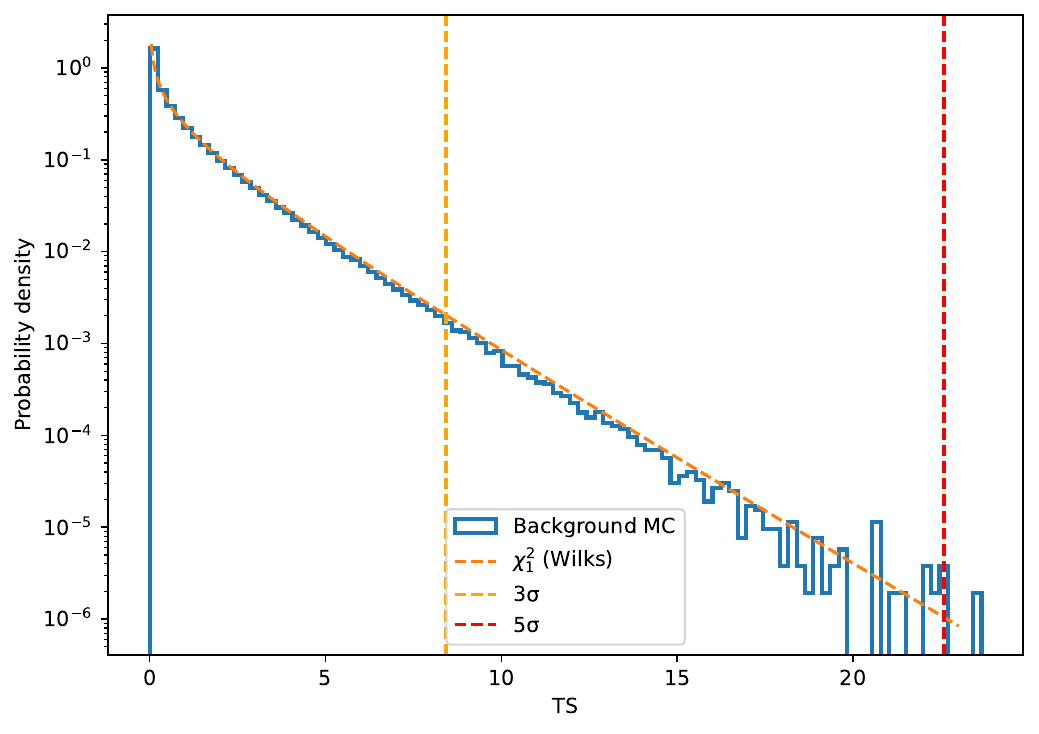} 
    \caption{
    Test statistic (TS) distribution obtained from $5\times 10^6$ background-only Monte Carlo simulations for the HUNT detector.The histogram shows the empirical TS distribution from simulations.The dashed black line indicates the theoretical expectation from Wilks' theorem for one free parameter ($\chi^2_1$). The vertical dashed orange and red lines indicate the TS thresholds corresponding to 3$\sigma$ and 5$\sigma$ single-tail significance, respectively.}
    \label{fig:ts_bg_distribution}
\end{figure}

To validate the use of $\sqrt{\mathrm{TS}}$ as an approximate significance estimator,we generate $5\times10^6$ background-only Monte Carlo realizations for the HUNT detector configuration.For each realization, the test statistic (TS) is computed in the same way as for the observed data.

Figure~\ref{fig:ts_bg_distribution} shows the resulting TS distribution. The empirical distribution is compared to the $\chi^2$ distribution with one degree of freedom, which represents the theoretical expectation under Wilks' theorem \citep{wilks1938,Cowan2011} for a single free parameter.The vertical dashed lines indicate the TS values corresponding to 3$\sigma$ (orange) and 5$\sigma$ (red) significance levels based on the empirical distribution.

As seen in the figure, the background TS distribution is in good agreement with the Wilks expectation for the TS range of interest,justifying the common approximation of statistical significance as $\sigma \simeq \sqrt{\mathrm{TS}}$. Small deviations may occur at very low TS values, but they do not affect the interpretation of the results in the parameter space considered.

{
\section{Compatibility with X-ray-bright Seyfert Stacking Results} \label{app:bat_stacking}

We validate our neutrino-generation framework by performing Monte Carlo simulations using northern-sky Swift/BAT Seyfert AGNs, ranked and weighted by the intrinsic 2--10~keV X-ray fluxes provided by the BAT AGN Spectroscopic Survey \citep[BASS;][]{BATagn2017ApJS}. In this test, the mock neutrino sky is generated from the full northern BAT AGN population, with the source weights proportional to the intrinsic 2--10 keV X-ray flux. A fraction of the total neutrino flux is attributed to the BAT AGN catalog, while the remaining events are modeled as an isotropic background, including both atmospheric and unresolved astrophysical neutrinos. The stacking catalog used in the likelihood analysis is then fixed a priori using only X-ray information, by selecting the brightest BAT AGNs.

\begin{figure}
\centering
\includegraphics[width=0.5\textwidth]{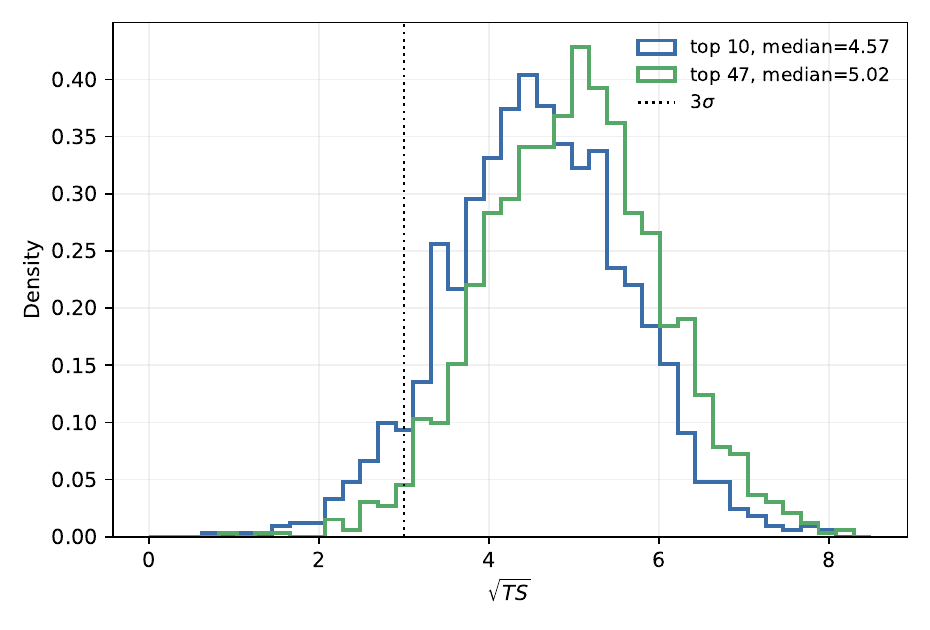}
\caption{
Distribution of the stacking test statistic \(\sqrt{TS}\) obtained from Monte Carlo realizations of the northern Swift-BAT AGN population.
In each realization, neutrino events are generated from the full BAT AGN catalog with weights proportional to the intrinsic 2--10 keV X-ray flux, while the stacking catalog is fixed a priori using only X-ray information.
The blue and green histograms correspond to stacking the top 10 and top 47 brightest BAT AGNs, respectively.
The vertical dotted line marks the \(3\sigma\) level.
The median significances are \(\sqrt{TS}=4.57\) and \(5.02\) for the top 10 and top 47 catalogs, respectively.
}
\label{fig:bat_prior_stacking}
\end{figure}

This procedure differs from selecting sources according to their neutrino excess and therefore avoids a posteriori source selection. It is intended as a qualitative consistency check rather than a reproduction of the official IceCube likelihood. As shown in Fig.~\ref{fig:bat_prior_stacking}, X-ray-bright BAT subcatalogs naturally yield few-sigma stacking signals in our benchmark setup. The median values are \(\sqrt{TS}=4.57\) for the top 10 catalog and \(\sqrt{TS}=5.02\) for the top 47 catalog. This demonstrates that bright-source stacking signals can be accommodated within the same population-based neutrino-generation framework, while the main focus of this work remains the detectability of the cumulative diffuse AGN component with future neutrino telescopes.
}

\bibliography{sample7}{}
\bibliographystyle{aasjournalv7}

\end{document}